\documentclass{article}

\usepackage{arxiv}

\usepackage[utf8]{inputenc} 
\usepackage[T1]{fontenc}   
\usepackage{hyperref}      
\usepackage{url}           
\usepackage{booktabs}     
\usepackage{amsfonts}      
\usepackage{nicefrac}      
\usepackage{microtype}   
\usepackage{lipsum}	
\usepackage{graphicx}
\usepackage[numbers]{natbib}
\usepackage{doi}

\title{LLC Intra-set Write Balancing}

\author{{Keshav Krishna} \\
	Department of Computer Science\\
	IIT Ropar\\
	Rupnagar, Punjab, 140001 \\
	\texttt{keshaviitropar@gmail.com} \\
	\And
	{Ayush Verma} \\
	Department of Computer Science\\
	IIT Ropar\\
	Rupnagar, Punjab, 140001 \\
	\texttt{} \\
}

\renewcommand{\shorttitle}{LLC Write Balancing}

\hypersetup{
pdftitle={LLC Intra-set Write Balancing},
pdfsubject={LLC Write Balancing},
pdfauthor={Keshav Krishna, Ayush Verma},
pdfkeywords={LLC, NVM, Write Endurance, Write Balancing, Interset, Variance, PC},
}

\begin{document}
\maketitle

\begin{abstract}
The increasing use of Non-Volatile Memory (NVM) in computer architecture has brought about new challenges, one of which is the write endurance problem. Frequent writes to a particular cache cell in NVM can lead to degradation of the memory cell and reduce its lifespan. To solve this problem, we propose a sample-based blocking technique for the Last Level Cache (LLC). Our approach involves defining a threshold value and sampling a subset of cache sets. If the number of writes to a way in a sampled set exceeds the threshold, the way is blocked, and writes are redirected to other ways. We also maintain a history structure to record the number of writes in a set and a PC-Table to use for blocking in unsampled sets. Based on blocking on sampled sets, variance of values stored in history is used to determine whether blocking had a positive impact or not, and on this basis, value corresponding to instruction pointer is incremented or decremented. This value is later used for blocking in unsampled sets. Our results show that our approach significantly balances write traffic to the cache and improves the overall lifespan of the memory cells while having better performance to the base-line system. Our approach can also be applied to other cache hierarchies and NVM technologies to mitigate the problem of write endurance.
\end{abstract}


\section{Introduction}
A type of computer memory known as NVMs (Non-Volatile Memories) can continue to store data even after the power is turned off. NVMs use a variety of technologies, including Phase Change Memory (PCM), Resistive Random Access Memory (RRAM) Fe-FET and MRAM, to store data persistently in contrast to conventional volatile memory such as DRAM, which requires constant power to maintain data retention \cite{si2021overview}. 

Phase Change Memory (PCM) is a type of NVM that uses phase change materials to store information in their amorphous and crystalline phases, which can be reversibly switched by the application of an external voltage \cite{raoux2014phase}. Resistive Random Access Memory (RRAM) is another type of NVM that has shown promise due to its high speed, low cost, enhanced storage density, potential applications in various fields, and excellent scalability \cite{zahoor2020resistive}. 

This is in contrast to conventional SRAM technology used in caches. SRAM (static random access memory) provides high performance and write endurance, but SRAM has low density and high leakage energy which leads to increased energy consumption and temperature of the chips \cite{7429336}. 

Because they can store data even when there is no power, NVMs are a highly relevant component for caches. This prevents the cache from needing lengthy warm-up times or pricey copying operations to restore its state following a power outage or system reboot.

NVMs provide a number of benefits, but they also have some drawbacks. Their low write endurance, or the number of times a memory cell may be programmed or wiped before it loses reliability, is a serious drawback \cite{drawbacks}, \cite{equalwrites}. Applying wear-leveling techniques, which uniformly spread write operations over the memory, can solve this problem but complicates the system.\cite{drawbacks} Another drawback is that NVM write speeds are slower than those of conventional SRAM caches, which might impact system performance.

NVMs also experience intra-set and inter-set write variance. While intra-set variation refers to fluctuation inside a set, inter-set variation refers to the change in write latency between various cache sets. These variances make it difficult to build effective caching techniques and can cause performance reduction.
To address this issue of intra-set write variation, we propose a new system that balances the write variability among the ways of the set while maintaining better performance than earlier approaches. We propose a variance based feed- back mechanism on some sampled sets to decide whether blocking had a positive impact or not, and using this feedback, do blocking on other sets.

We ran our system against previous approaches and received better performance, while still tackling the write endurance issue.

Many intra-set cache wear leveling techniques have been proposed in the existing literature. This section briefly discusses all such techniques. 

\cite{wang20132} described i2wap, a wear levelling technology that reduces both inter-set and intra-set wear levelling. Probabilistic Set Line Flush (PoLF) is the first technique for reducing intra-set write variance. Swap Shift, which adjusts the mapping of two sets after a defined number of writes called Swap Threshold, is the second strategy for reducing inter-set write variation.

\cite{jokar2015sequoia} reports on another wear levelling approach named Sequoia for reducing writing variance. The inter-set write variance is decreased in the first strategy, G-OAS, by partitioning the cache into numerous groups of sets and modifying the set mapping of heavily written and weakly written sets in each group while taking the counter associated with each set into account. WAD decreases intra-set write variation by shifting data inside the set once the corresponding set counter (RSC) saturates.

The EqualChance algorithm described by \cite{mittal2014equalchance} decreases intra-set write variation by performing a transfer/swap operation within the cache set. \cite{qureshi2009enhancing} proposes a technique called Start-Gap that uses the two registers Start and Gap to move the line location to the neighbour place after a specific number of writes. \cite{kotra2016re} reports on a ReRAM NUCA that handles the lifetime problem in a performance aware manner.

LastingNVCache, as proposed by \cite{mittal2014lastingnvcache}, decreases intra-set write variation by including write counters with each cache block. When the counter hits a certain threshold, the write process is aborted by invalidating the block. A word level strategy presented by \cite{wang2017word} decreases writing variance by exploring data with a small width. A hybrid cache architecture is as follows: Ayush, as presented by \cite{mittal2015ayush}, decreases write variation by moving data blocks between cache zones.

EqualWrites, as proposed by \cite{equalwrites}, decreases write variation by shifting the block from a write-intensive point inside the set to a chilly location. Several write throttling techniques at various levels of memory hierarchy have already been presented. A Lady technique described by \cite{lee2014effective} extends the life of flash-based solid-state devices by regulating writes dynamically according on workload parameters. Mellow writes, as described by\cite{zhang2016mellow} enhances main memory longevity by introducing a wear limitation approach. To improve endurance, slow writes are performed with lesser power dissipation. 

In this work, we compare our proposed approach with \cite{equalwrites}.

The paper is organized as follows: Introduction and Related works are discussed in Section 1. Methodology is reported in Section 2. Description of our algorithm is presented in Section 3. Section 4 lists the results and analysis. Finally, we conclude this paper in Section 4, followed by acknowledgement in Section 5.

\section{Methodology}

\subsection{Motivation}
The motivation behind this is that there is a large variation of writes among ways of a set. This variance can be attributable to a number of variables, including production differences, transistor ageing, and environmental fluctuations, which can have an impact on the performance and dependability of the cache in different ways.


The write endurance of the cache—the number of write operations that can be made on the cache before it fails—can be significantly impacted by this write variance. Some cache ways may receive much more writes than others in a cache with high write variation, which could cause those ways to break early and reduce the cache's total write endurance. This may lead to higher maintenance and replacement expenses as well as decreased system reliability and performance.

The impacts of write variance in NVM-based caches have been addressed using a variety of strategies, including dynamic write balancing and wear levelling. These strategies try to evenly distribute write operations among the cache paths, lowering the risk of early failure and boosting the cache's write endurance.

These strategies proposed so far only consider most recent writes among the sets and then use thresholds to block certain ways. But This may lead to excessive blocking and creation of some undesired effects like oscillatory nature of writes, which is explained in detail in later section. 

So we propose a feedback based algorithm that considers the effects of previous blocking before making a decision.

\subsection{Key Idea}
Large write variation between ways of a set can occur in the final level cache, which reduces the cache's write endurance. An algorithm has been created to address this issue, and it is based on the notion that by sampling a small number of sets, we can also make accurate blocking judgements on larger sets. This means that we can predict the behaviour of other sets in the cache by studying the behaviour of a few selected sets.

The method takes use of the fact that programmes and data structures with identical access patterns are often comparable. The programme can identify which approaches are most and least frequently utilised by analysing the behaviour of a small number of sets.

The programme also takes advantage of another key idea: instruction pointer bias by noting that only a small number of instruction pointers are frequently processed in the context of blocking methods in the last level cache to address the write endurance problem. Similar sets and methods of accessing these instruction pointers, as well as similar writing across different executions, characterise them.

This observation is used by the algorithm to reinforce certain instruction pointers positively or negatively. The approach grants a specific instruction pointer positive reinforcement if blocking by that pointer has a positive impact on the cache in sampled sets. In contrast, the approach reinforces a given instruction pointer negatively if blocking by that port has a negative impact on the cache in sampled sets.

The decision to block a path in other sets is then made using the value assigned to each instruction pointer based on whether it received positive or negative reinforcement. In other words, the algorithm uses the correlation between the cache access patterns and the instruction pointer bias to determine the appropriate blocking decisions for other sets. The approach ensures that the last level cache has a balanced write endurance across all ways by only blocking those ways that cause substantial write variations.

\section{Algorithm Description}

\subsection{Counters}
A structure is made to store the count of write operations done in that particular block of cache and the instruction pointer for that write. This structure is a 2D array, the array dimension being the number of sets and ways of the cache to be simulated. Interval size was divided based on number of cycles executed by the processor(s). It can correctly measure time elapsed irrespective of other factors such as CPU core count. In the counter 2D array, count of write operations(counter values) performed in that i-th interval are stored. After the start of every cycle, these counter values and instruction pointer values are set to be zero. The value of the counter is incremented when handlefill() operation i.e. request for write in that empty block of cache or data from main memory is brought into cache is called for that particular set and way of the cache, and the instruction pointer that caused that write is updated as the instruction pointer value.

\begin{figure}
	\centering
	\includegraphics[width=15cm,height=7cm,angle=0]{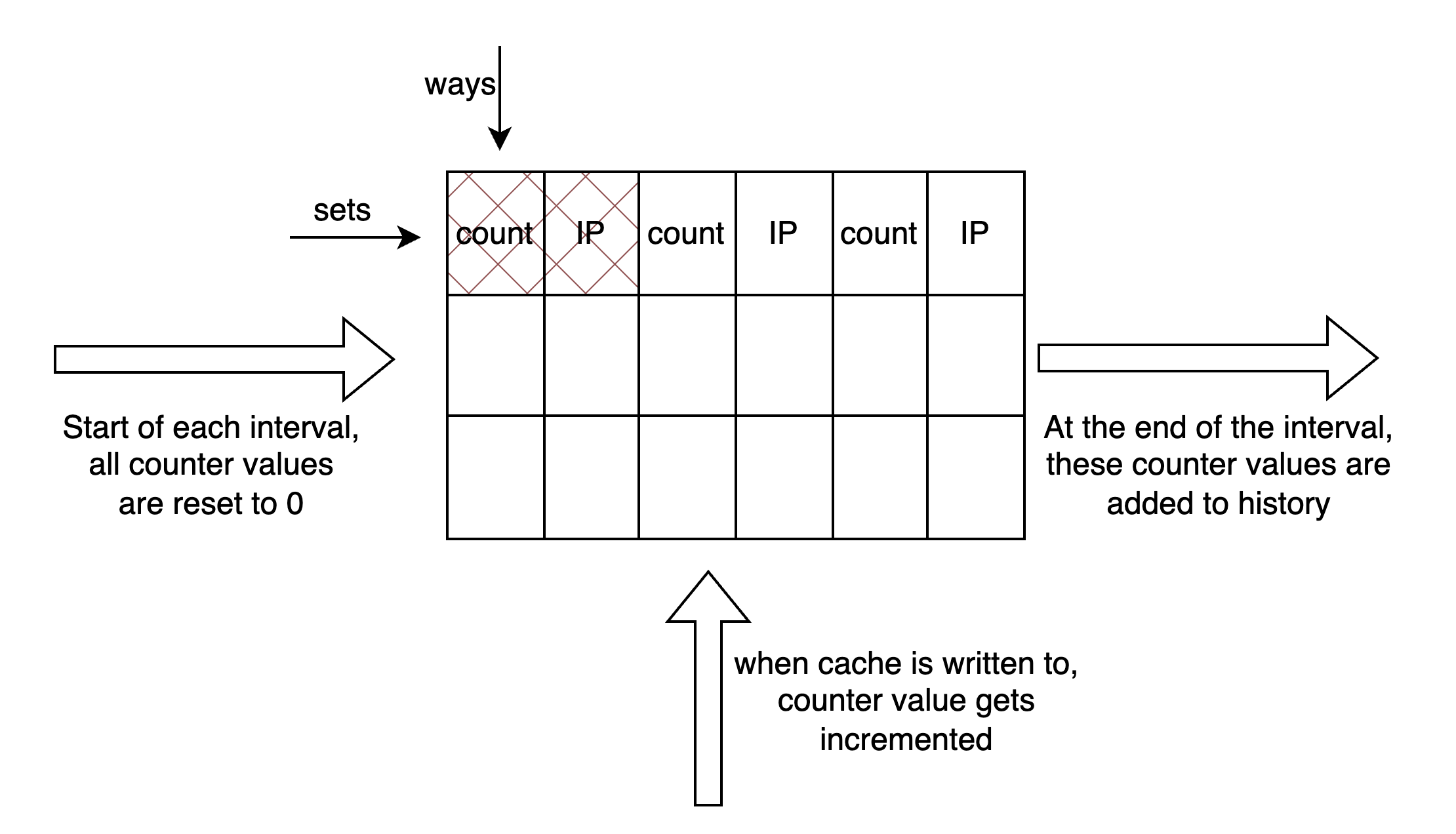}
        \caption{Structure of counter used to record number of writes in I cycles and the instruction pointer that caused the write}
	\label{fig:fig1}
\end{figure}

\subsection{History storing structure}
To extend upon the counter values, which were stored for all round operation of the cache, and did not give insight on how writes were happening after every ‘I’ cycles, we made a history keeping structure. The structure is a map of key and vector of vector of values, which stored for each set, what are the counter values for previous k ‘I’ cycles(the number of writes and instruction pointer). After every I cycles, the counter values were added for each set into the map, and if size of map for that set is greater than k(set to be 8), most ancient counter values were removed, so that the size of the structure for each set index is limited to k vectors. The history structure is only maintained for LLC(last level cache).

\begin{figure}
	\centering
	\includegraphics[width=15cm,height=7cm,angle=0]{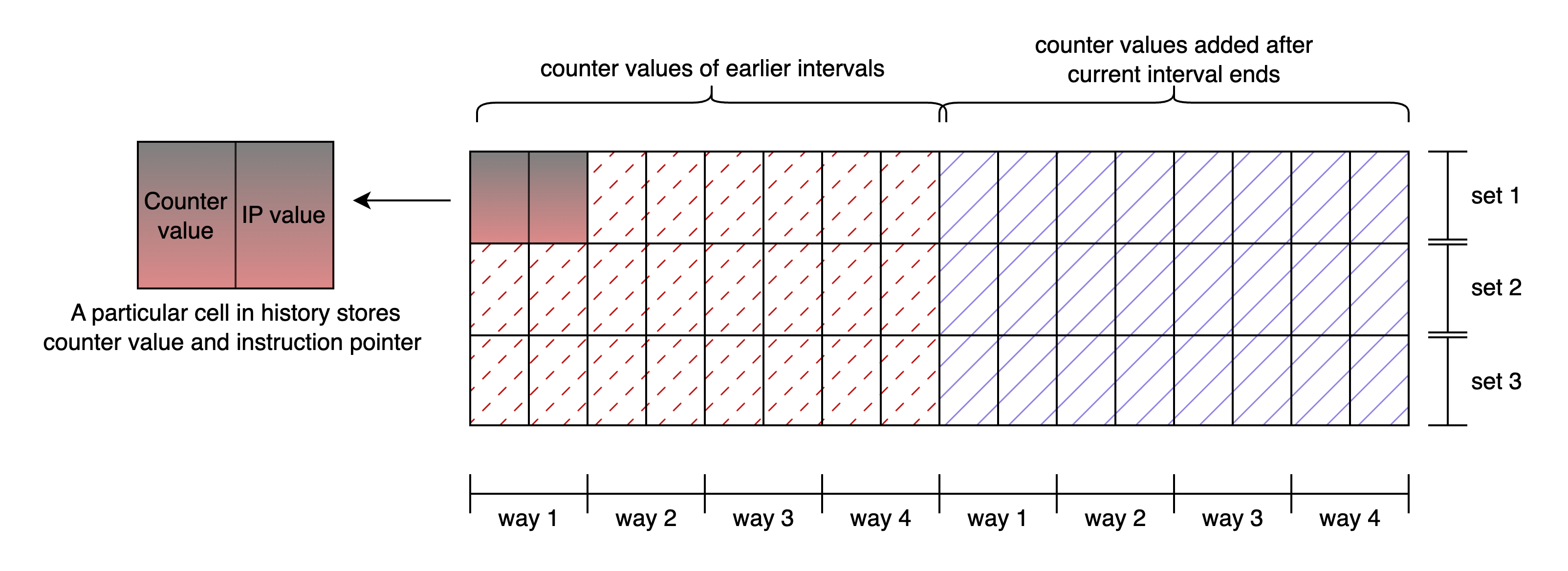}
        \caption{Structure of history which stores past 'k' counter values for each set. For implementation of algorithm, k = 8 is taken.}
	\label{fig:fig2}
\end{figure}

\subsection{Sampled sets and Unsampled sets}
Two types of sets are implemented in LLC: Sampled sets and Non-sampled sets. Sampled sets are a subset of all the sets in the cache, and they are decided based on mathematical function using bitmasking. Out of 2048 total sets, 32 sets are taken to be sampled sets, and the rest are Non-Sampled sets. Counters and previous 'k' history storing structure is made only for Sampled sets. This is because storing counters and history structure for every sets will increase the hardware overhead largely and deployment of this algorithm on an actual LLC would not be feasible. Having only 32 sampled sets to store the counter and history will be very hardware friendly. On the other hand, non-sampled sets are the remaining sets in the cache that are not part of the sampled sets. These sets do not have counters and history structures. 

The sampled sets are selected based on the helper function, which compares the first 6 bits of the set index with the last 6 bits of the set index shifted by the logarithm of the cache set size. This ensures that the sampled sets are distributed evenly across the cache while also reducing the chances of collisions between different cores accessing the same sets. 

\subsection{PC Table}
An important observation which is 'Instruction Pointer bias' is utilized in the algorithm. It was observed that only a limited number of instruction pointers were frequently processed in the context of blocking methods. These frequently accessed instruction pointers exhibited similar access patterns and methods, as well as similar writing behavior across different program executions. As a result, by taking advantage of this bias, it is feasible to optimize write operations and reduce total cache wear, resulting in balanced write endurance. 

PC Table is maintained for every Instruction pointer. This table is a map for 'uint64' to 'int' that maps the instruction pointer to its respective value. This value is set to be zero initially and is incremented or decremented based on whether blocking a particular cache cell of the sampled set had a positive impact or negative impact on the remaining cache cells in the next iteration. This PC-Table is accessed to decide whether blocking is to be done or not in Non-Sampled sets depending on the value corresponding to that particular instruction pointer.  

\subsection{Variance based feedback} 
After blocking a particular cache cell of a sampled set, based on comparison of the counter values with the threshold value, its impact on remaining cache cells(ways) of the same set is observed in the next iteration. The variance among the counter values of sets are calculated in the next iteration. The Instruction Pointer that blocked the particular cell is trained based on the variance calculated in the next iteration. The variances due to blocking by this particular instruction pointer in previous 'k' cycles is also taken into consideration. Weighted mean of these variances is then calculated, giving more weights to recent variances. Based on the value of this weighted mean of variances, feedback to increase or decrease the value corresponding to that instruction pointer is done in the PC-Table. 

\begin{figure}
    \advance\leftskip-1cm
    \begin{minipage}{\linewidth}
        \includegraphics[width=17cm,height=9cm,angle=0]{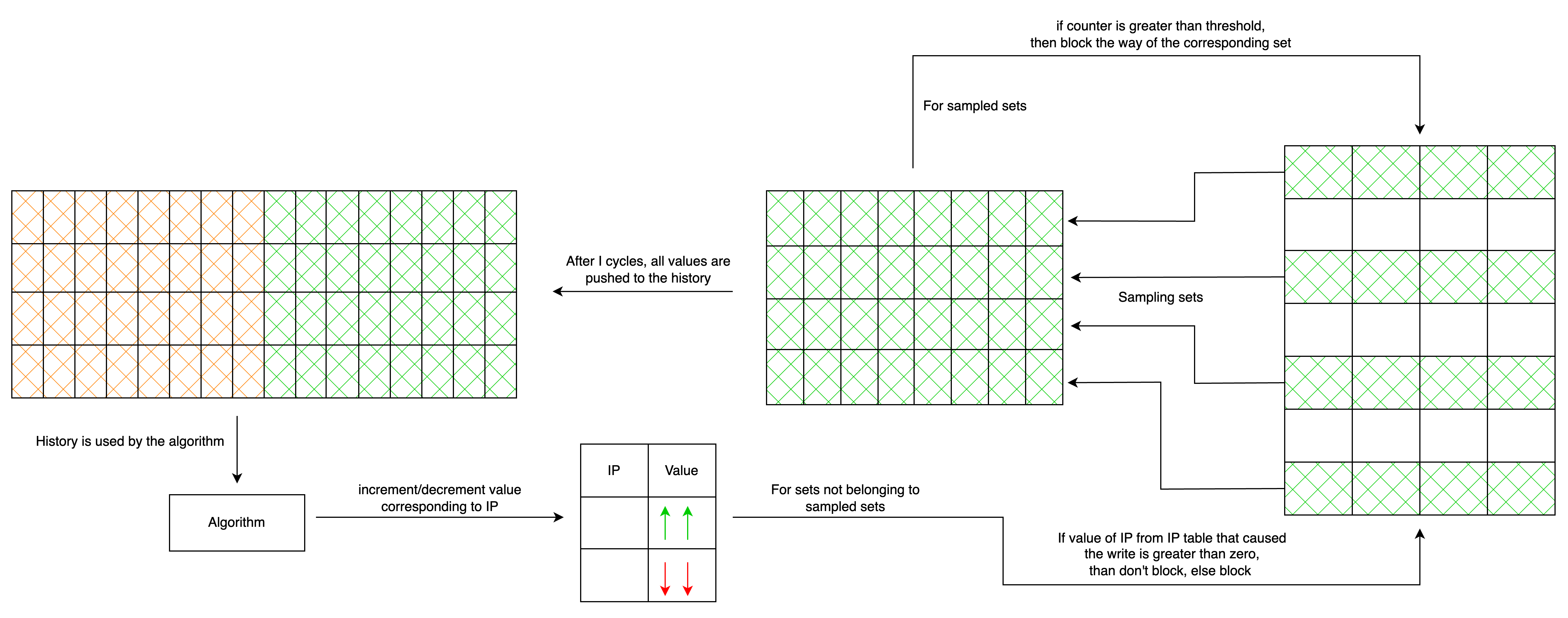}
        \caption{Schematic diagram of flow of the algorithm. Some sets are sampled as sample sets, and for them counter values are calculated and stored in history after every 'I' cycles. At the end of each of the interval of 'I' cycles, PC-table is trained based on whether blocking due to an instruction pointer had positive effect or not. For unsampled sets, whenever a write occurs, the instruction pointer is used to access the PC-table and get the value. If value < 0, then that way is blocked for write operations, else normal write can occur.}
    \end{minipage}
\end{figure}

\subsection{Blocking}
Blocking of a particular cache cell is handled differently on the basis whether that cell belongs to one of the Sampled sets or Non-Sampled sets. If the set to which cache belongs to Sampled set, at the start of current cycle, its previous cycle counter value is compared with threshold. If the respective counter value is less than threshold value, no blocking is done and the working would proceed normally. On the other hand, if counter value exceed theshold, that particular cache cell is blocked. The way returned by chechhit() function is set to be -1(analogous to write-miss), implying that some other unblocked cell needs to be returned by the replacement function. The replacement algorithm (srrip) also takes care of the fact that the victim cell returned by it should not be a blocked cell. In the next iteration, the effect of this blocking is observed by calculating variance. The Instruction Pointer that blocked the particular cell is trained based on the variance calculated in the next iteration. Weighted mean of the variance in previous 'k' cycles due to blocking by instruction pointer is calculated. Based on the value of this weighted mean of variances, feedback to increase or decrease the value corresponding to that instruction pointer is done in the PC-Table.  

If the cache cell belongs to Non-Sampled sets, the instruction pointer that brought this particular cell is noted. The value corresponding to this instruction pointer is fetched from the PC-Table. If this value is negative, cache cell is blocked in current iteration, else it is unblocked. Instruction pointer biasness described in section 3.3.4 is the basis for this decision.

\begin{table}
	\caption{Simulation parameters used for Last Level Cache}
	\centering
	\begin{tabular}{lll}
		\toprule
		\cmidrule(r){1-2}
		LLC Parameter     & Values \\
		\midrule
		No. of sets  & 2048     \\
		No. of sampled sets & 32      \\
		No. of non-sampled sets & up to 2016  \\
            No. of ways & 16 \\
            Threshold & 29 \\
		\bottomrule
	\end{tabular}
	\label{tab:table}
\end{table}

\section{Results}
\subsection{Problems with existing approaches}

\begin{figure}
	\centering
	\includegraphics[width=15cm,height=10cm,angle=0]{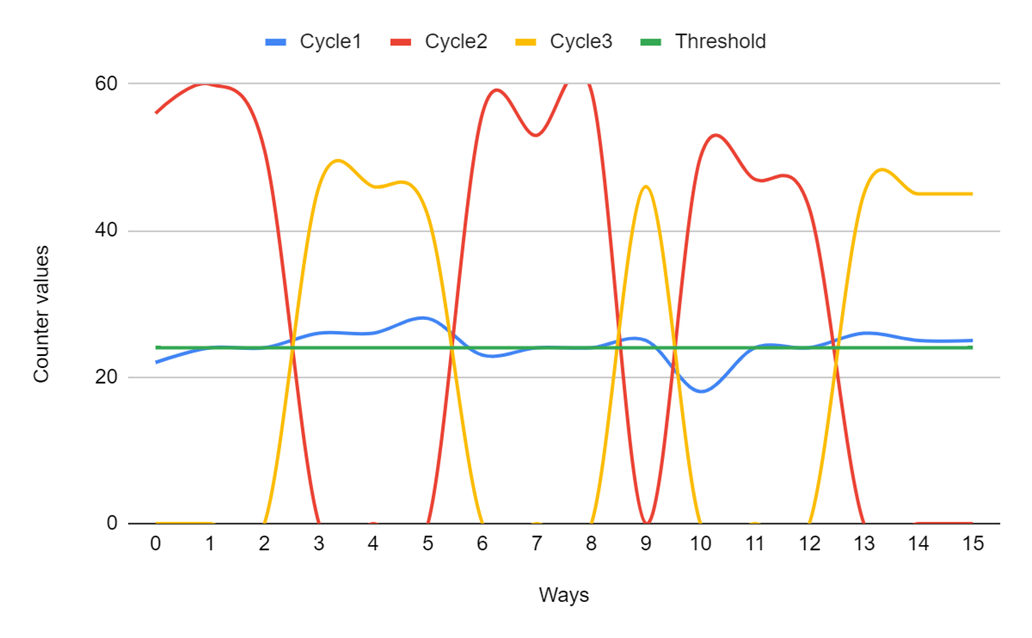}
        \caption{Graph of counter values vs ways. It depicts how blocking using equal writes or other similar approaches can lead to an oscillatory nature.}
	\label{fig:fig3}
\end{figure}
The graph \ref{fig:fig3} shows the variation of counter values in different cycles for a particular set and its corresponding ways if only threshold based blocking mechanism is used. The green line is the threshold value and blue, red and yellow lines depict cycle-1, cycle-2 and cycle-3 respectively. Oscillatory nature of counter values can be very well observed from the graph i.e. in the previous cycle, the way for which the threshold was exceeded is blocked for the current cycle. But, the write counter values for those unblocked ways have now exceeded the threshold and so they will get block in the next cycle.

\subsection{IP Bias}

\begin{figure}
	\centering
	\includegraphics[width=15cm,height=10cm,angle=0]{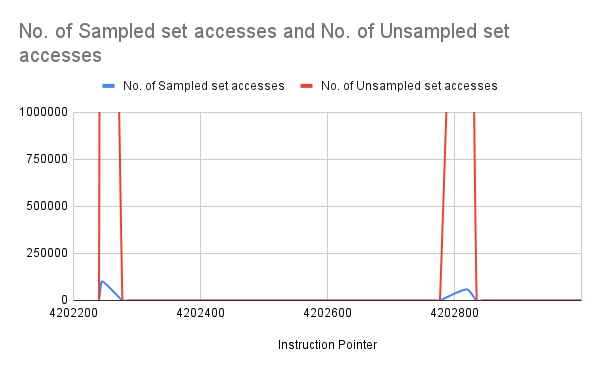}
        \caption{Graph of number of accesses vs instruction pointers for sampled sets and not sampled sets. We can observe that some sets instruction pointers have large accesses and others very less.}
	\label{fig:fig4}
\end{figure}

The graph \ref{fig:fig4} depicts the Instruction Pointer Bias which has been utilised in our algorithm. The blue line depicts number of sampled set accesses and red line depicts number of unsampled set accesses by an Instruction Pointer. It is worth noting here that only 32 sampled sets are present out of total 2048 total sets. Only a limited number of instruction pointers were frequently processed in the context of blocking methods. These frequently accessed instruction pointers exhibited similar access patterns and methods, as well as similar writing behavior across different program executions. As a result, by taking advantage of this bias, it is feasible to optimize write operations and reduce total cache wear, resulting in balanced write endurance

\subsection{Variance}

\begin{figure}
	\centering
	\includegraphics[width=15cm,height=10cm,angle=0]{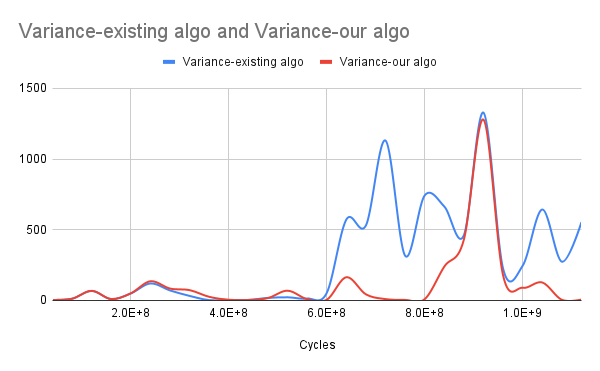}
        \caption{Graph of variance vs number of cycles elapsed for existing algorithm and proposed algorithm. We can see variances are almost always less for our proposed algorithm.}
	\label{fig:fig5}
\end{figure}

The graph\ref{fig:fig5} compares the variances of ways of a particular set in difference cycles obtained by the existing approach(blue line graph) to those obtained by our algorithm(red line graph). For lower number of cycles, both the methods have similar variances. But for larger number of cycles, our method gives lower variance value among the ways of the set implying that appropriate levelling of write operation is handled by our algorithm.

\subsection{IPC}

\begin{figure}
	\centering
	\includegraphics[width=15cm,height=10cm,angle=0]{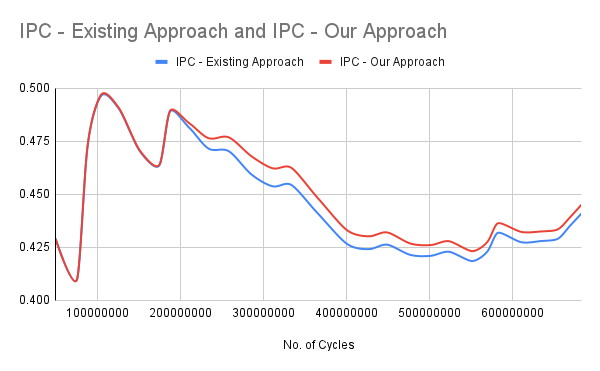}
        \caption{Graph of IPC vs number of cycles. We can see that our proposed algorithm has better IPC than existing approaches.}
	\label{fig:fig6}
\end{figure}

The graph\ref{fig:fig6} shows a comparison of the IPC(Instructions per Cycle) value of existing approach (\cite{equalwrites}) and our approach. When number of cycles is low, both the approaches have similar IPC values. As number of cycles increases, our approach gives better IPC value as compared to existing approach. IPC value is a great metric to compare performance of different algorithms applied to cache-architecture. Note that the above graph is drawn for \emph{libquantum} trace.

\subsection{IPC trends over different traces}

\begin{table}
	\caption{IPC values of existing approaches and proposed algorithm for different traces.}
	\centering
	\begin{tabular}{lll}
		\toprule
		\cmidrule(r){1-2}
		Traces & IPC (Existing Approach) & IPC (Our Approach) \\
		\midrule
		gcc\_13B.trace.xz & 0.123613 & 0.123836     \\
		bwaves\_1861B.trace.xz & 0.437062 & 0.437127      \\
		mcf\_158B.trace.xz & 0.0376413  & 0.0376554  \\
            libquantum\_964B.trace.xz & 0.277188  & 0.277591 \\
		\bottomrule
	\end{tabular}
	\label{tab:table2}
\end{table}

IPC values obtained by existing approach and our approach over different traces has been given in the table\ref{tab:table2}. For all the four different traces, IPC value of our approach is higher than the existing method. This means more number of instructions are executed in same number of cycles meaning that our approach has better performance than existing method (\cite{equalwrites}).

\subsection{Miss ratio over different traces}
\begin{table}
	\caption{Miss ratio (misses/total accesses) of existing approaches like equal writes against our proposed algorithm for different traces.}
	\centering
	\begin{tabular}{lll}
		\toprule
		\cmidrule(r){1-2}
		Traces & Miss Ratio(Existing Approaches) & Miss Ratio(Our Approach) \\
		\midrule
		gcc\_13B.trace.xz & 0.930459 & 0.92914     \\
		bwaves\_1861B.trace.xz & 0.983 & 0.98299      \\
		mcf\_158B.trace.xz & 0.98379  & 0.98347  \\
            libquantum\_964B.trace.xz & 0.89084  & 0.89084 \\
		\bottomrule
	\end{tabular}
	\label{tab:table3}
\end{table}

Miss Ratio which is the number of misses divided by total cache accesses is another performance metric in cache architecture. Less the miss ratio, better is the performance. LLC miss ratio value for different traces has been given in the table\ref{tab:table3} each for existing approach and our approach. For \emph{libquantum trace}, both approach have the same miss ratio. While for other three traces, our approach has less miss ratio than the existing approach.

\section{Conclusion}
NVM(Non-Volatile Memory) has major advantages over SRAM, like its non-volatile, low power consumption, and high density. But to make it usable in caches, a major challenge of NVM, namely its write endurance issue, needs to be overcome.
The main reason for this write endurance issue is that different ways of a set as well as different sets, do not have the same number of writes. This inter-set and intra-set write variability causes some blocks to fail, hampering the working of the cache.
To overcome this issue, many existing approaches were proposed. A common feature of these approaches is that they block, that is, make a way not accessible for writes. This then directs writes to a cold way, balancing the writes.
But a common problem with these approaches is that as they only use previous writes to make the decision to block or not, they suffer from problems associated with excessive blocking, like an oscillatory nature of the number of writes.
In this paper, we have proposed a feedback-based blocking method that uses previous 'k' histories to make the decision to block a way or not. We compared our proposed system with an existing approach(\cite{equalwrites}) and found our algorithm has better IPC(Instructions per cycle). Also, the proposed system's variance was better, implying better balancing of writes among the ways.
Thus if we can solve the write endurance problem, we can better utilize NVMs.

\section{Acknowledgement}
I would want to express my sincere thanks and gratitude to everyone who helped bring this project to a conclusion.

First and foremost, I'd want to express my gratitude to my project supervisors, Dr. Shirshendu Das and Dr. Sudeepta Mishra, for their excellent guidance and support during the project. Their knowledge and suggestions have been very much valuable in influencing the project's direction and ensuring that it fulfills the requisite requirements.

I would also want to thank Mr. Prathamesh, the project's teaching assistant, for his assistance. His advice and support have been vital in the project's successful conclusion.

\bibliographystyle{unsrtnat}
\bibliography{references}

\end{document}